\documentclass[aps,twocolumn,showpacs]{revtex4}

\usepackage[xdvi]{graphicx}
\usepackage{latexsym}
\usepackage{amsbsy}

\begin{document}

\date{\today}

\title{Domain wall mobility in nanowires: transverse versus vortex walls}

\author{R.\ Wieser, U.\ Nowak and K.\ D.\ Usadel}
\affiliation{Theoretische Tieftemperaturphysik,
  Gerhard-Mercator-Universit\"{a}t Duisburg, 47048 Duisburg, Germany}

\begin{abstract}
  The motion of domain walls in ferromagnetic, cylindrical nanowires
  is investigated numerically by solving the Landau-Lifshitz-Gilbert
  equation for a classical spin model in which energy contributions
  from exchange, crystalline anisotropy, dipole-dipole interaction,
  and a driving magnetic field are considered.  Depending on the
  diameter, either transverse domain walls or vortex walls are found.
  The transverse domain wall is observed for diameters smaller than
  the exchange length of the given material. Here, the system behaves
  effectively one-dimensional and the domain wall mobility agrees with
  a result derived for a one-dimensional wall by Slonczewski. For low
  damping the domain wall mobility decreases with decreasing damping
  constant.  With increasing diameter, a crossover to a vortex wall
  sets in which enhances the domain wall mobility drastically.  For a
  vortex wall the domain wall mobility is described by the
  Walker-formula, with a domain wall width depending on the diameter
  of the wire. The main difference is the dependence on damping: for a
  vortex wall the domain wall mobility can be drastically increased
  for small values of the damping constant up to a factor of
  $1/\alpha^2$.
\end{abstract}

\pacs{75.10.Hk, 75.40.Mg, 75.60.Ch}
\maketitle

Arrays of magnetic nanowires are possible candidates for patterned
magnetic storage media \cite{rossPRB00,nielschJMMM02}. For these
nanowires and also for other future magneto-electronic devices the
understanding of domain wall motion and mobility is important for the
controlled switching of the nanostructure. In a recent experiment the
velocity of a domain wall in a NiFe/Cu/NiFe trilayer was investigated
using the GMR effect \cite{onoSCIENCE99}.  The measured velocities
were compared with the Landau-Lifshitz formula for domain wall motion
\cite{landauPZS35}. This comparison was used to determine the damping
constant of the trilayer, a quantity which is usually not known a
priori.  However, several formulas for the velocity of a domain wall
can be found in the literature
\cite{landauPZS35,malozemoffBOOK79,schryerJAP74,dillonBOOK63,garaninPA91b}
which are derived in different limits and all in (quasi)
one-dimensional models neglecting the possible influence of
non-uniform spin structures within the domain wall. Thus the question
arises in how far these formulas are applicable to real three
dimensional domain structures. To shed some light onto this problem
we numerically investigate the domain wall mobility in nanowires
starting from a three dimensional local spin model.

In the following we consider a classical spin model with energy
contributions from exchange, crystalline anisotropy, dipole-dipole
interaction, and a driving magnetic field. Such a spin model for the
description of magnetic nanostructures \cite{nowakARCP01} can be
justified following different lines: on the one hand it is the
classical limit of a quantum mechanical, localized spin model, on the
other hand it might be interpreted as the discretized version of a
micromagnetic continuum model, where the charge distribution for a
single cell of the discretized lattice is approximated by a point
dipole.  For certain magnetic systems their description in terms of a
lattice of magnetic moments may even be based on the mesoscopic
structure of the material, especially when a particulate medium is
described.

However, our intention is not to describe a particular material but to
investigate a general model Hamiltonian which is
\begin{eqnarray}
  \label{H}
  {\cal H} = &-& J \sum\limits_{\langle ij \rangle} {\mathbf S}_i
    \cdot {\mathbf S}_j 
    - \mu_s {\mathbf B} \cdot \sum\limits_i {\mathbf S}_i
    - D_e \sum\limits_i (S_i^z)^2 \nonumber \\
   & - & \omega \sum\limits_{i<j} \frac{3({\mathbf S}_i \cdot {\mathbf
      e}_{ij})({\mathbf e}_{ij} \cdot {\mathbf S}_j) - {\mathbf S}_i
    \cdot {\mathbf S}_j}{r^3_{ij}},
\end{eqnarray}
where the ${\mathbf S}_i = {\boldsymbol \mu}_i/\mu_s$ are three
dimensional magnetic moments of unit length on a cubic lattice.

The first sum is the ferromagnetic exchange between nearest neighbors
with coupling constant $J$.  The second sum is the coupling of the
spins to an external magnetic field $B$, the third sum represents a
uniaxial anisotropy, here, with $D_e > 0$, favoring the $z$ axis as
easy axis of the system, and the last sum is the dipolar interaction
where $w = \mu_0 \mu_s^2 /(4 \pi a^3)$ describes the strength of the
dipole-dipole interaction.  The ${\mathbf e}_{ij}$ are unit vectors
pointing from lattice site $i$ to $j$ and $r_{ij}$ is the distance
between these lattice sites in units of the lattice constant $a$.

The underlying equation of motion for magnetic moments which we
consider in the following is the Landau-Lifshitz-Gilbert (LLG)
equation,
\begin{equation}
\frac{\partial {\mathbf S}_i}{\partial t} =
  -   \frac{\gamma}{(1+\alpha^2)\mu_s}
  {\mathbf S}_i \times \Big[{\mathbf H}_i(t) +
  \alpha \big ({\mathbf S}_i \times {\mathbf H}_i(t) \big) \Big],
\label{e:llg}
\end{equation}
with the gyromagnetic ratio $\gamma = 1.76 \times 10^{11} (\mathrm{
  Ts})^{-1}$, the dimensionless damping constant $\alpha$ (after
Gilbert), and the internal field $ {\mathbf H}_i(t) = - \partial {\cal
  H} /\partial {\mathbf S}_i$.
  
We simulate cylindrical systems being parallel to the $z$-axis with a
typical length of 256 lattice sites and different diameters $d$. Due
to shape as well as crystalline anisotropy the equilibrium
magnetization is aligned with the long axis of the system. However, we
start the simulation with an abrupt, head-to-head domain wall as
initial configuration, letting the wall relax until a stable state is
reached. The distance of the initial wall position from the end is
approximately 1/3 of the system length. Then we switch on the driving
magnetic field $B$ along the easy axis and wait until a stationary
state is reached for some time interval in which the velocity $v$ of
the wall is constant while the wall is moving through the central part
of the wire. We calculate the domain wall velocity from the
magnetization versus time data, averaged over a period of time where
no influence of the finite system size on the domain wall can be
observed, i. e. until the wall approaches the other end of the wire.

\begin{figure}[h]
  \includegraphics[height=8cm]{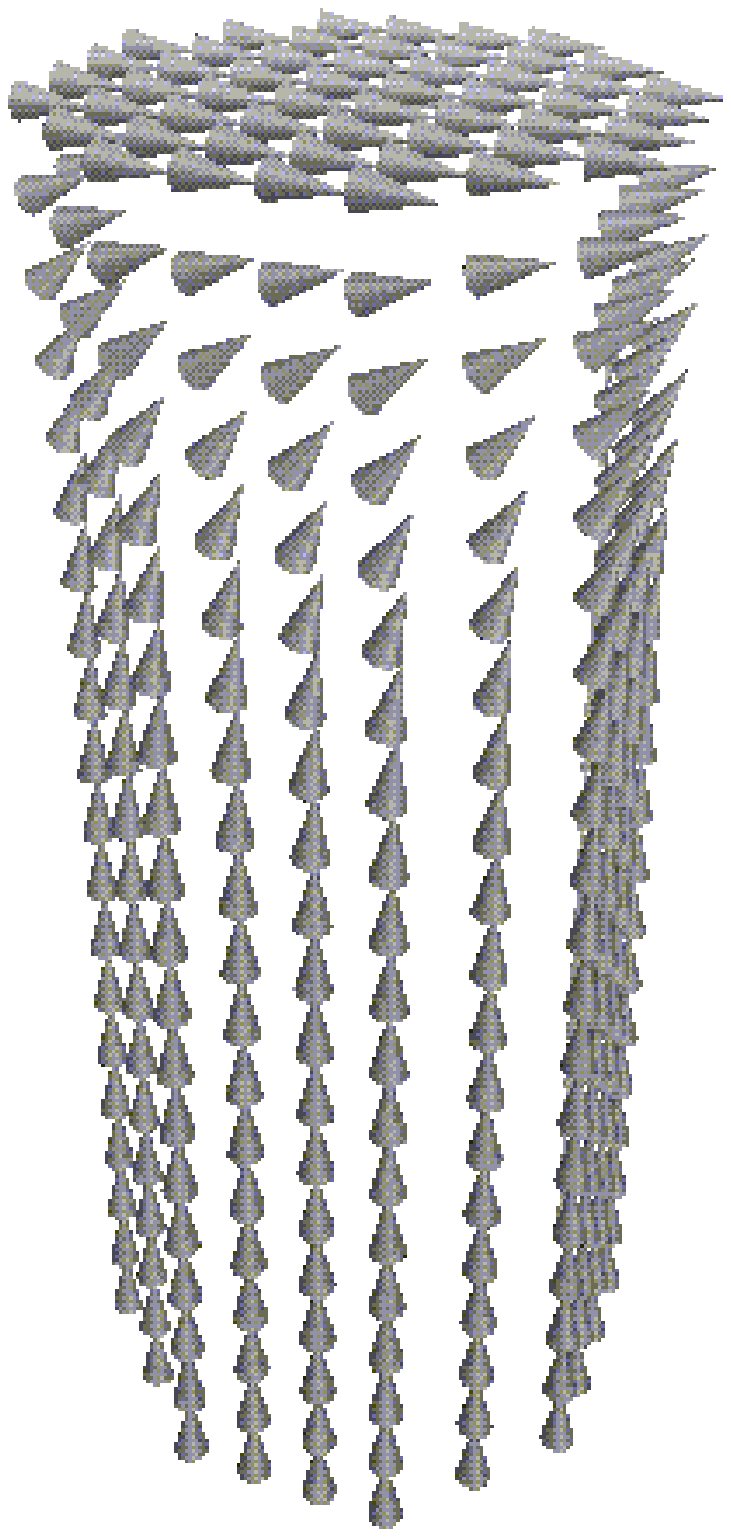}
  \includegraphics[height=8cm]{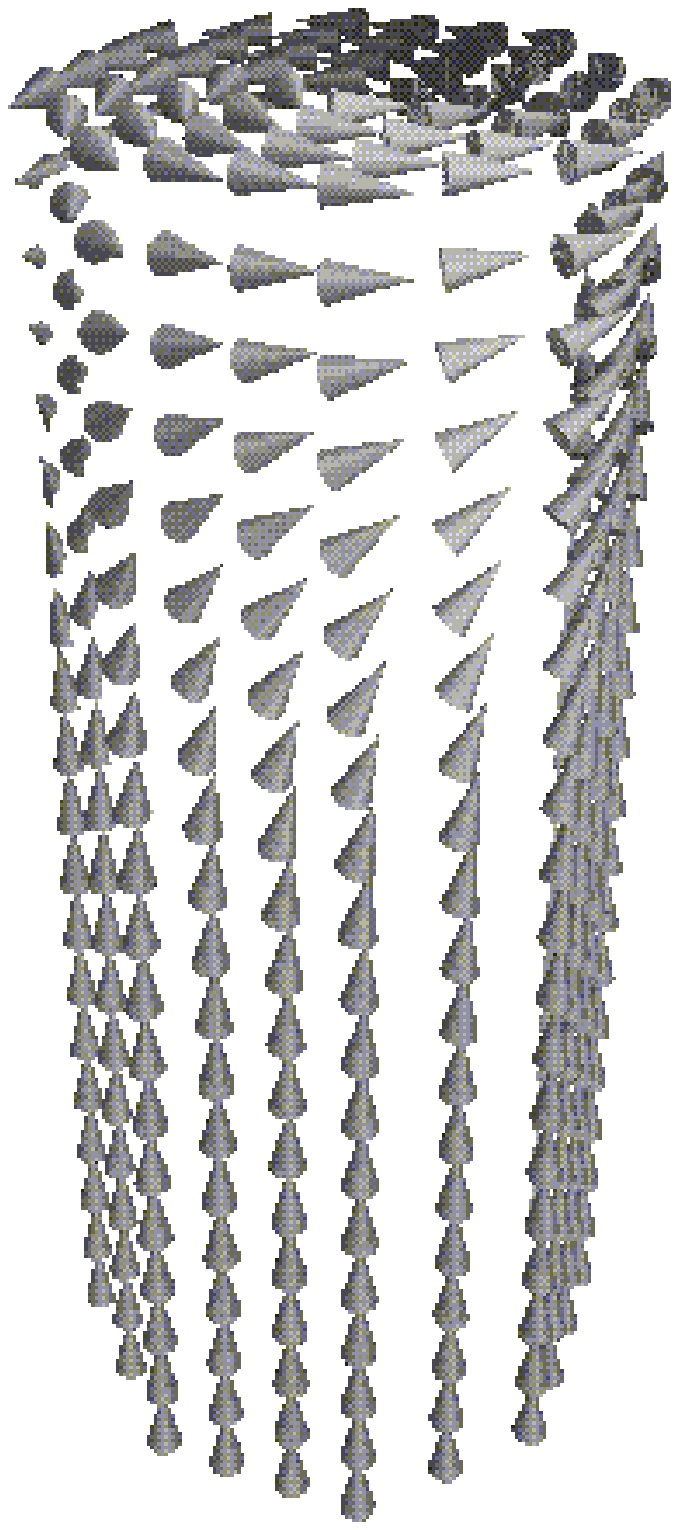} \caption{Snapshots of a
    transverse (left, $\omega/J = 0.003$) and a vortex (right,
    $\omega/J = 0.2$) wall.  The diameter $d=8$ is kept constant while
    the exchange length of the system is varied. Shown is only a part
    of the system below the current wall position. $D_e/J = 0.05$.}
  \label{f:snaps}
\end{figure}

Inspection of the stationary state of the moving domain wall shows
that, depending on the ratio $\omega/J$, either transverse domain
walls or vortex walls are found. Representative spin configurations
are shown in Fig. \ref{f:snaps}. The transverse domain wall (left hand
side) is observed for diameters smaller than the exchange length
$d_{ex}/a = \pi \sqrt{J/(6 \omega \zeta(3))}$ of the system
\cite{hinzkeJMMM00} where $\zeta(3) \approx 1.2$ is Riemann's
Zeta-function (see also \cite{hubertBOOK98} for the exchange length in
continuum theory where $3 \zeta(3)$ is replaced by $\pi$).  Here all
spins within cross-sectional planes perpendicular to the wire axis are
parallel so that the system is effectively one dimensional. Note, that
the spin precession leads to a rotation of the spin direction within
the wall while it is moving.

With increasing dipolar interaction, a crossover to a vortex wall sets
in (right hand side of Fig. \ref{f:snaps}) which is now energetically
favorable since the vortex structure leads to a flux closure. These
findings are in agreement with corresponding spin model simulations of
thermally activated reversal \cite{hinzkeJMMM00} and micromagnetic
results \cite{forsterJAP02,hertelJMMM02} obtained from simulations of
the LLG equation using a micromagnetic continuum model.

\begin{figure}[h]
  \includegraphics[width=8cm]{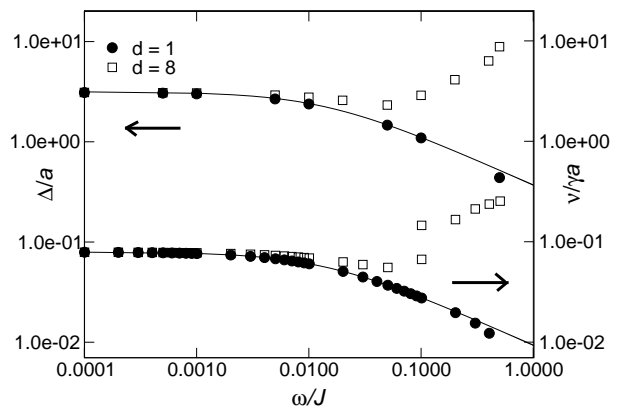}
  \caption{Domain wall velocity and domain wall width versus dipolar
    coupling for a spin chain and a cylindrical system, respectively.
    $D_e/J = 0.05$, $\mu_s B/J = 0.05$, $\alpha = 1$. The solid lines
    correspond to Eqs.  \ref{e:wall} and \ref{e:slon}, error bars are
    smaller than the symbol size.}
  \label{f:vpropd}
\end{figure}

In the following we turn to the investigation of the influence of the
domain wall width and structure on its velocity.  Fig. \ref{f:vpropd}
compares the dependence of the domain wall width $\Delta$ and domain
wall velocity $v$ on the strength of the dipolar coupling.  Here, the
domain wall width was determined numerically by fitting an
$\tanh$-profile to the easy axis magnetization of the moving wall in
the stationary state where the magnetization is averaged over
cross-sectional planes. However, it should be mentioned that for large
dipolar interaction in a vortex wall the wall profile cannot
accurately be described by a simple tanh-profile. Note, that even in
the limit $\omega \to 0$ the wall is stabilized by the additional
crystalline anisotropy $D_e$.

For a spin chain ($d = 1$) the domain wall is necessarily always
planar while for the system with larger diameter a crossover to a
vortex wall occurs. The crossover can be identified as a jump of the
domain wall velocity for the $d=8$ data. Fig.  \ref{f:vpropd}
demonstrates that for a transverse wall the domain wall velocity is
proportional to the wall width. For a vortex wall this is at least
qualitatively the case. Width and velocity of transverse walls
decrease with increasing dipolar interaction while for vortex walls
the opposite is true. The crossover itself leads to a jump of the wall
velocity not the wall width.  \vspace{7mm}

For a transverse wall the domain wall velocity is well described by an
equation derived by Slonczewski as the lower limit for one-dimensional
domain wall motion \cite{malozemoffBOOK79},
\begin{equation}                                
  v  = \frac{\gamma}{\alpha + 1/\alpha} \Delta_{\mathrm B} B.
  \label{e:slon}
\end{equation}
Here, $\Delta_B$ is the well-known Bloch wall width
\begin{equation}
  \Delta_{\mathrm B} = a \sqrt{\frac{J}{2 \left(D_e + 3 \omega
        \zeta(3)\right)}},
  \label{e:wall}
\end{equation}
where, for our case, the denominator $D_e + 3 \omega \zeta(3)$
estimates the effective anisotropy coming from shape as well as
crystalline contributions (as before in a continuum theory $3
\zeta(3)$ is replaced by $\pi$). Both equations above are drawn in
Fig. \ref{f:vpropd} as solid lines and they agree very well with the
numerical data for transverse walls.

\begin{figure}[h]
  \includegraphics[width=8cm]{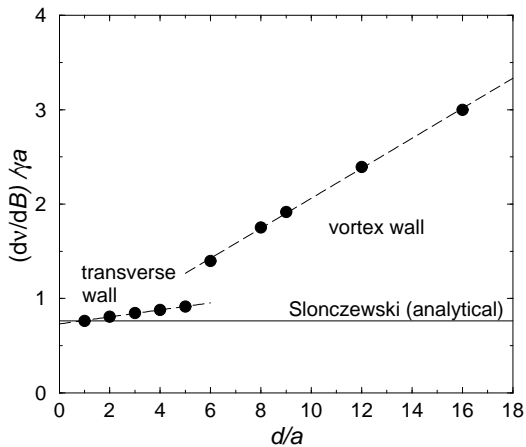}
  \caption{Domain wall mobility versus diameter of cylindrical
    systems. The model parameters are $\omega/J = 0.003$, $D_e/J =
    0.1$, $\alpha = 1$. The constant line corresponds to Eq.
    \ref{e:slon}, the dashed lines are guides to the eye, error bars
    are smaller than the symbol size.}
\label{f:mobility}
\end{figure}

In the following we focus on the mobility of vortex walls.  The
crossover from transverse to vortex wall can also be observed while
varying the diameter of the system keeping the exchange length
constant. Since for sufficiently small driving fields the domain wall
velocity is proportional to the field in Fig. \ref{f:mobility} we
directly show the domain wall mobility ${\mathrm d} v / {\mathrm d} B$
versus diameter of the system.  Obviously there are two distinct
regions with distinct wall mobility behavior. For low diameters where
the transverse wall is found the system behaves effectively
one-dimensional and in the limit $d \to 1$ the domain wall mobility
follows Eq. \ref{e:slon}.  With increasing diameter the observed width
of the transverse domain wall increases little due to the dipolar
interaction leading to small deviations from the analytic Slonczewski
result assuming a Bloch wall width. Nevertheless, we confirmed
numerically, that Eq. \ref{e:slon} is still valid when the Bloch wall
width $\Delta_{\mathrm B}$ is replaced by the actual (numerically
determined) width of the transverse wall.

Increasing the diameter of the system a crossover from transverse to
vortex wall is observed with a drastic increase of the domain wall
mobility. As seen in Fig. \ref{f:vpropd}, the reason for this effect is not
a comparably drastic change of the wall width.  Instead, as we will
discuss in the following for a vortex wall the domain wall mobility
follows a law with a different dependence on the damping constant,
namely the Landau-Lifshitz formula \cite{landauPZS35},
\begin{equation}                                
  \frac{{\mathrm d} v}{{\mathrm d} B} = \frac{\gamma}{\alpha} \Delta,
  \label{e:ll}
\end{equation}
where in our case $\Delta$ is the actual domain wall width of the
vortex wall.  The equation above is a limit of the more general Walker
equation \cite{schryerJAP74,dillonBOOK63,garaninPA91b},
\begin{equation} 
  v = \frac{\gamma B a}{\alpha}\sqrt{\frac{J}{2(D_e + D_h
      \sin^2\phi)}},
  \label{e:walker}
\end{equation}
which was derived for sufficiently small driving fields for a system
with an additional hard-axis anisotropy $D_h$. This anisotropy forces
the equilibrium magnetization into an easy plane. Walkers formula is
valid as long as the spin motion takes place in one plane which is
defined by a constant angle $\phi$ to the easy plane of the system.
$\phi$ is given as
\begin{equation}
  \sin \phi \cos \phi = \frac{\mu_s B}{\alpha 2 D_h},
  \label{e:phi}
\end{equation}
where this equation also defines a condition for the validity of the
walker formula. For a given $\alpha$ there exists a maximum field
value (or vice versa for a given field a minimum $\alpha$ value)
beyond which the spin motion is no longer restricted to one plane and
instead an irregular precessional motion starts \cite{schryerJAP74}.
Note, that the Landau-Lifshitz formula is the $\phi = 0$ limit of the
Walker equation, i. e. the limit of a strong hard axis anisotropy
which forces the spin motion into the easy plane.

\begin{figure}[h]
  \includegraphics[width=5cm]{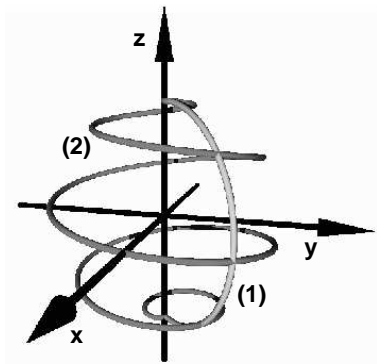}
  \caption{Two different paths for the reversal of a spin. While (2)
    is dominated by precession, as in a transverse wall, the path (1) is
    restricted to one plane as in a vortex wall.}
  \label{f:paths}
\end{figure}

The equations above were derived for one dimensional systems and the
question arises why these formula should be valid for the motion of a
vortex wall with a non-uniform spin structure in cross-sectional
planes. For a qualitative understanding we note that the motion of the
spins within each spin chain which is parallel to the wire axis is
indeed restricted to a certain plane passing through the spins
positions.  For a spin chain at the surface of the cylinder and in the
limit of small driving fields these are tangential planes of the
cylinder surface. The responsible force which keeps the spin motion of
each chain in this plane is for a vortex wall not a hard axis
anisotropy --- as in the original calculation --- but the energetical
principle which forms the vortex, i. e., the combination of exchange
and dipolar interaction.  Since this is the condition under which
Walkers formula was derived it seems to be plausible that Eq.
\ref{e:walker} describes the wall mobility in the case of an extended
spin system as long as the spin motion during the reversal takes
place in one plane. For a transverse wall, on the other hand, the
situation is different: the precession of the wall leads to the fact
that the motion of each single spin consists of precession and
relaxation with no restriction to one single plane.  These two
different paths for the reversal of a spin are sketched in Fig.
\ref{f:paths}.

\begin{figure}[h]
  \includegraphics[width=8cm]{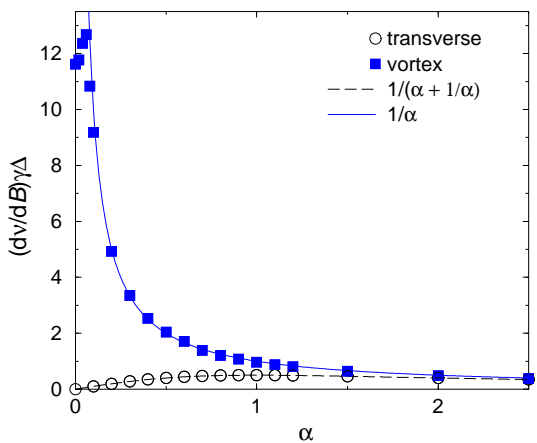}
  \caption{Reduced domain wall mobility versus damping constant. The model
    parameters are $\omega/J = 0.01$ (transverse wall) and $\omega/J =
    0.7$ (vortex wall) respectively, $D_e/J = 0.05$, $\mu_s B/J =
    0.05$, $d = 4$. Error bars are smaller than the symbol size.}
  \label{f:damping}
\end{figure}

The main difference between Eqs. \ref{e:slon} and \ref{e:ll} is the
dependence on damping. This is demonstrated in Fig. \ref{f:damping}
which shows the ratio of the calculated wall mobility and the
numerically determined domain wall width for two different strengths
of dipolar interaction leading to the two different wall shapes.  In
the high damping limit both formulas agree. For a transverse wall the
mobility shows a maximum at $\alpha = 1$ and for lower damping the
domain wall mobility decreases with decreasing damping constant.  In
the limit $\alpha \to 0$ only a precession of the domain wall remains
without an effective wall motion along the wire.

For a vortex wall the domain wall mobility increases with decreasing
damping constant following a $1/\alpha$ law as long as one is above a
critical value $\alpha_c$.  As was discussed in connection with Eq.
\ref{e:phi} this value $\alpha_c$ sets the limit of pure relaxational
spin motion.  As was discussed before, the role of the hard axis
anisotropy $D_h$ in Eq. \ref{e:phi} in our case is played by the
combination of exchange and dipolar interaction which forms the vortex
and forces the spin motion into one plane. We would like to stress
that for experimental systems the low damping limit is more relevant.
Here, the difference between the two domain wall mobilities (reduced
to the domain wall width) can be extremely large, up to a factor of
$1/\alpha^2$.

For smaller values of $\alpha$ below the critical one the mobility
decreases again and finally converges to a finite value since even for
$\alpha = 0$ the wall can move. In this limit the LLG equation
conserves the energy of the system, and lowering the Zeeman energy
leads to an increase of exchange energy, leaving an excited spin
system behind the wall.  These observations are also in agreement with
the calculations of Walker \cite{schryerJAP74}.

\begin{figure}[h]
  \includegraphics[bb = 18 40 515 340, width=8cm]{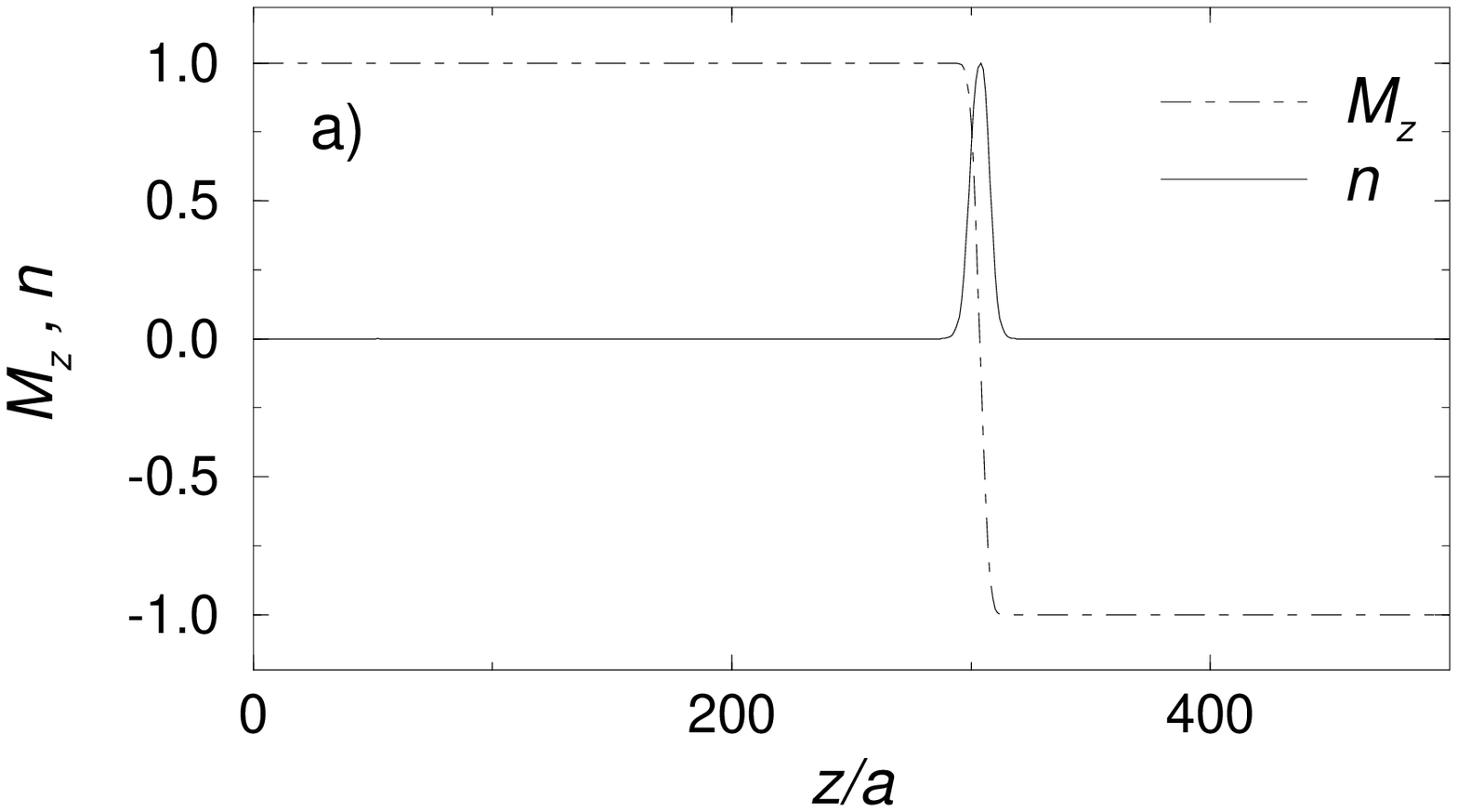}\\
  \includegraphics[bb = 18 40 515 340, width=8cm]{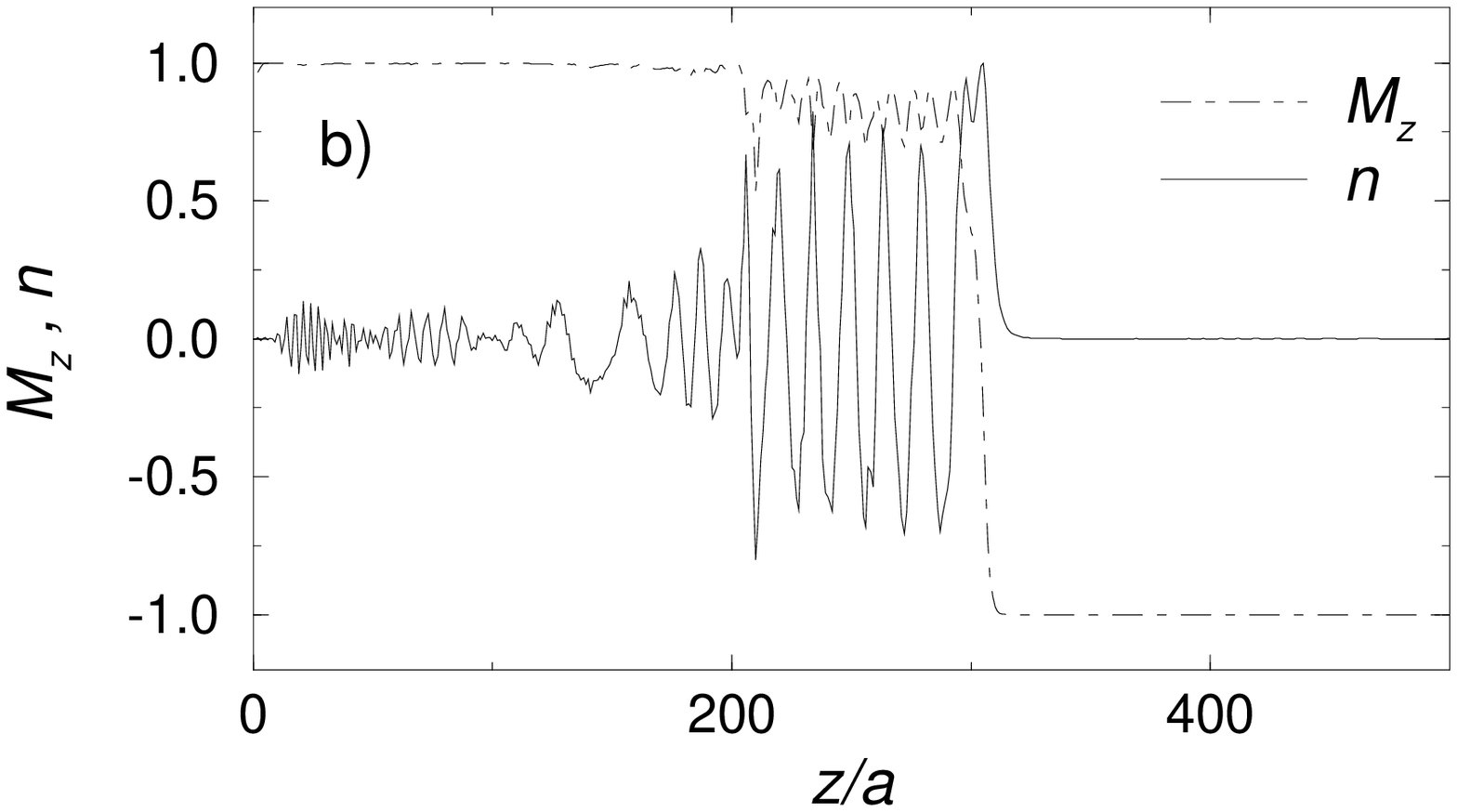}
  \caption{Profiles and winding numbers of moving vortex 
    walls in a) the high damping limit ($\alpha = 1$) and b) the low
    damping limit ($\alpha = 0$). $d = 8$, $\omega/J = 0.2$, $D_e/J =
    0.05$, $\mu_s B/J = 0.1$}
  \label{f:winding}
\end{figure}

This effect is demonstrated in Fig. \ref{f:winding}. Here, profiles of
the moving walls are shown as well as the so-called winding number
\[
  n = \frac{1}{2 \pi R} \int ({\rm rot}\underline S)_zdxdy
\]
which is calculated numerically over all perpendicular planes of the
wire. The winding number is a measure for the existence of vortices.
$n=1$ means that all spins along the boundary are aligned building a
ring with perfect flux closure.  Fig.  \ref{f:winding} a) shows the
high damping limit with a perfect vortex in the center of the wall. In
the very low-damping limit, Fig.  \ref{f:winding} b) the situation is
much more complicated. Here, behind the moving wall (smaller $z$) an
excited spin system is left with vortex-type spin waves which are
ejected from the moving wall.

To conclude, in agreement with prior work
\cite{hinzkeJMMM00,forsterJAP02,hertelJMMM02} we have found different
wall structures for driven domain walls in cylindrical systems,
transverse and vortex walls, depending on the diameter of the system
as compared to the exchange length of the given material. While for
vortex walls the domain wall velocity is described by the formula from
Walker, transverse walls follow a formula from Slonczewski. In both
cases the domain wall velocity is proportional to the domain wall
width. The main difference is the dependence on the damping constant.
For small values of the damping constant this difference can lead to
drastic differences where the velocity of the vortex wall is up to a
factor of $1/\alpha^2$ larger.  The reason for this difference is
probably the fact that each spins motion in the case of the vortex
wall is completely within one single plane as it is the case for the
model where the Walker formula was derived for, while this is not the
case for a transverse wall where the precession of the wall leads to a
three dimensional spin motion.

\section*{Acknowledgments}
The authors thank D. Garanin and S. L\"{u}beck for helpful
discussions.  This work has been supported by the Deutsche
Forschungsgemeinschaft (SFB 491 and NO290).


\begin{thebibliography}{10}

\bibitem{rossPRB00}
C.~A. Ross, R.~W. Chantrell, M. Hwang, M. Farhoud, T.~A. Savas, Y. Hao, H.~I.
  Smith, F.~M. Ross, M. Redjdal, and F.~B. Humphrey, Phys. Rev. B {\bf 62},
  14252  (2000).

\bibitem{nielschJMMM02}
K. Nielsch, R.~B. Wehrspohn, J. Barthel, J. Kirschner, S.~F. Fischer, H.
  Kronm\"uller, T. Schweinb\"ock, D. Weiss, and U. G\"osele, J. Magn. Magn.
  Mat. {\bf 249},  234  (2002).

\bibitem{onoSCIENCE99}
T. Ono, H. Miyajima, K. Shigeto, K. Mibu, N. Hosoito, and T. Shinjo, Science
  {\bf 284},  468  (1999).

\bibitem{landauPZS35}
D.~L. Landau and E. Lifshitz, Phys.~Z.~Sowjetunion {\bf 8},  153  (1935).

\bibitem{malozemoffBOOK79}
A.~P. Malozemoff and J.~C. Slonczewski, {\em Magnetic Domain Walls in Bubble
  Materials} (Academic Press, New York, 1979).

\bibitem{schryerJAP74}
{N.\,L.~Schryer} and {L.\,R.~Walker}, J. Appl. Phys. {\bf 45},  5406  (1974).

\bibitem{dillonBOOK63}
J.~F. Dillon,  in {\em Magnetism}, edited by G.~T. Rado and H. Suhl (Academic
  Press, New York, 1963), Vol.~1, p.\ 149.

\bibitem{garaninPA91b}
{D.\,A.~Garanin}, Physica A {\bf 178},  467  (1991).

\bibitem{nowakARCP01}
U. Nowak,  in {\em Annual Reviews of Computational Physics IX}, edited by D.
  Stauffer (World Scientific, Singapore, 2001), p.\ 105.

\bibitem{hinzkeJMMM00}
D. Hinzke and U. Nowak, J. Magn. Magn. Mat. {\bf 221},  365  (2000).

\bibitem{hubertBOOK98}
A. Hubert and R. Sch\"{a}fer, {\em Magnetic Domains} (Springer-Verlag, Berlin,
  1998).

\bibitem{forsterJAP02}
H. Forster, T. Schrefl, D. Suess, W. Scholz, V. Tsiantos, R. Dittrich, and J.
  Fidler, J. Appl. Phys. {\bf 91},  6914  (2002).

\bibitem{hertelJMMM02}
R. Hertel, J. Magn. Magn. Mat. {\bf 249},  251  (2002).

\end{thebibliography}
\end{document}